\documentclass[superscriptaddress,secnumarabic,
amssymb,amsmath,nobibnotes,aps,prd,showkeys,showpacs,nofootinbib]{revtex4}%
\usepackage{graphicx}
\usepackage{epsf}
\usepackage{bm}
\usepackage{amsmath}
\usepackage{amsfonts}
\usepackage{amssymb}
\usepackage{epstopdf}
\usepackage{natbib}
\usepackage{color}%
\usepackage[normalem]{ulem}
\usepackage{soul}
\providecommand{\U}[1]{\protect\rule{.1in}{.1in}}
\newcommand{\be}{\begin{equation}}
\newcommand{\ee}{\end{equation}}

\newcommand{\mincir}{\raise
-3.truept\hbox{\rlap{\hbox{$\sim$}}\raise4.truept\hbox{$<$}\ }}
\newcommand{\magcir}{\raise
-3.truept\hbox{\rlap{\hbox{$\sim$}}\raise4.truept\hbox{$>$}\ }}

\newtheorem{remark}{Remark}[section]

\begin{document}



\title{A simple inflationary quintessential model}



\author{Jaume de Haro\footnote{E-mail: jaime.haro@upc.edu}}
\affiliation{Departament de Matem\`atiques, Universitat Polit\`ecnica de Catalunya, Diagonal 647, 08028 Barcelona, Spain} \author{Jaume Amor\'os\footnote{E-mail: jaume.amoros@upc.edu}}
\affiliation{Departament de Matem\`atiques, Universitat Polit\`ecnica de Catalunya, Diagonal 647, 08028 Barcelona, Spain}
\author{Supriya Pan\footnote{E-mail: span@research.jdvu.ac.in}}
\affiliation{Department of Mathematics, Jadavpur University, Kolkata-700032, West Bengal, India}


\thispagestyle{empty}

\begin{abstract}
In the framework of a flat Friedmann-Lema{\^\i}tre-Robertson-Walker (FLRW) geometry, { we present a non-geodesically past complete model of our universe without the big bang
singularity at finite cosmic time,} describing its evolution starting from its early inflationary era up to the present accelerating phase. We found that a hydrodynamical fluid with
nonlinear equation of state could result in such scenario, which after the end of this inflationary stage, suffers a sudden phase transition and
enters into the stiff matter dominated era, and the universe becomes reheated due to a huge amount of particle production. Finally, it asymptotically enters into the
de Sitter phase concluding the present accelerated expansion. {{Using the reconstruction technique,} we also show that, this background
   provides an extremely simple  inflationary quintessential potential whose inflationary part  is given by the well-known $1$-dimensional Higgs potential, i.e., a
   {\it Double Well Inflationary} potential,  and the quintessential one by an exponential potential that leads to a deflationary regime after this inflation, and it can depict the current
   cosmic acceleration at late times. Moreover the Higgs potential
    leads to a power spectrum of the cosmological
perturbations which fit well with the latest Planck estimations}. Further, we compared our viable potential with some known inflationary quintessential
potential, which shows that our { quintessential
model, that is, the Higgs potential combined with the exponential one}, is an improved version of them
because it contains an analytic solution that allows us to perform all analytic calculations.
{Finally, we have shown that the introduction of a non zero cosmological constant simplifies the potential considerably with an analytic behavior
of the background which again permits us to evaluate all the quantities analytically.}

\end{abstract}

\vspace{0.5cm}


\maketitle

\section{ Introduction}

 The complete evolution of our universe is still a mystery, and probably, one of the most interesting topics in 
 the history of cosmology. Until now, we have some theories describing different phases of our universe, in agreement with the latest observations, which
 tell us that our universe underwent a rapid accelerating phase during its very early evolution, namely, the inflation \cite{Inflation,Ade}, and presently it is going
 through a phase of accelerated expansion \cite{Supernovae,Complimentary}. The gap between these two successive accelerating expansions is described by
 three sequential decelerated phases, the first one is the stiff matter dominated era, then there was a radiation dominated phase, and finally, before
 its current accelerating phase, the universe  was matter dominated. However, since the beginning of modern cosmology, the big bang still remains as one of
 the controversial issues for cosmologists. Hence, it has been questioned several times, and alternatively, an existence
of
some { kind of ``nonsingular'' universe (a model of our universe without  finite cosmic time singularity) \cite{emergent} has been proposed just to replace this big bang singularity,
but the evolution
of the universe will remain same. As a result, the prospect of a unified cosmological theory attracted cosmologists in order to describe the present accelerating phase while
still tracing back the early evolution of the universe. To find such a viable cosmological theory,
it has been shown that, {in the flat Friedmann-Lema{\^\i}tre-Robertson-Walker (FLRW) geometry}, the bulk viscous cosmology \cite{hp}
and the gravitationally induced ``adiabatic'' particle creations \cite{phpj} can provide some
quintessential potential, {which can describe a scenario of our universe with no} big bang singularity at finite cosmic time,} the early inflationary phase, and finally the current accelerating universe.
{Considering the same spacetime, in
the present work,} we have shown that this { same} feature with the said properties could also be understood in the context of pure General Relativity, if one assumes a simple
nonlinear Equation of State (EoS).
The idea of the model is very simple: for a large value of the Hubble parameter, namely $H$,  the dynamics of the system is depicted by a first order
linear differential equation, namely, the Raychaudhuri equation. With this choice, { in spite of the model leads to a geodesically past incomplete universe},
the big bang singularity can be removed { at finite cosmic time},
while preserving an inflationary regime during the early
evolution of the universe. At the end of this inflationary period, we introduce
an abrupt phase transition to a deflationary period, where the energy density of the background decays as $a^{-6}$,
{{}where $a$ is the scale factor of the FLRW universe}. Moreover, due to the phase transition, the energy density of the produced particles decreases as $a^{-4}$, and thus,
eventually becoming dominant, {{}it reheats the universe}, which matches with the standard hot Friedmann model. Only at very late times, the background comes back to dominate
going asymptotically to a de Sitter solution, which depicts the current cosmic acceleration.\newline

{
However, as we discuss at the beginning of section 3,
to study the origin of cosmological perturbations, an introduction of an  scalar field with its corresponding potential which could mimic the early inflationary dynamics
is mandatory, because it is not clear how hydrodynamical cosmological perturbations leads to a nearly flat
power spectrum of perturbations that match with current observational data. Surprisingly,
 we have found that the inflationary quintessential potential that  mimics this  dynamics is a simple and well-known potential. Effectively, we have obtained a combination
 of the $1$-dimensional Higgs potential -a  {\it Double Well Inflationary} potential \cite{DWI}- and an exponential one. The first one is  responsible for inflation while
 the other one for quintessence. This is for us, although indirectly, the main reason to deal with the background that we present in the work, which in the first instance has been
 obtained using a nonlinear EoS.

  Further, in order to check the viability { of} this inflationary potential, we calculate the associated parameters related to the power spectrum from the background
  perturbations, such as, the spectral index, its running, and the ratio of tensor to scalar perturbations. We found that the model fits well with the recent parameter
  estimations by Planck \cite{Ade}. Finally, we compared this scalar field potential with some quintessential models, and noticed that our model potential has some
  noteworthy points which differentiate it from the previously introduced potentials, and hence presents a viable potential alternative to the existing ones.}\newline

The paper is organized as follows: {{}In section \ref{model}, introducing the model with its nonlinear EoS, we solve the dynamical equations analytically and discuss its properties.}
{{}In section \ref{scalar-field}, we establish that the dynamics governed by the model} could be mimicked by a single scalar field whose { extremely simple}
potential is a combination of { the Higgs potential}, and an exponential one, {where we note that the incorporation} of a scalar field
is mandatory when one deals with the origin of cosmological perturbations.
Section { 4} is devoted to the study of cosmological perturbations showing that the theoretical results provided by our models fit well with current observed data \cite{Ade,Planck}.
The reheating process via gravitational particle production is studied in section { 5}, where we show that our model provides a reheating temperature below $10^9$ GeV, which is
required in order to have a successful nucleosynthesis.
{ Comparison with other inflationary quintessential models is done in section 6.
Finally, in {the} last section, we
introduce the cosmological constant, {and obtain} a very simple potential
which is the matching of the { Higgs} potential and the cosmological constant.
This potential also leads to a  background that can be calculated analytically and it allows us to
perform {all analytic calculations.}}\newline

The units used throughout the paper are $\hbar=c=8\pi G=1$.

\section{The model}
\label{model}
Our model {which is non-geodesically past complete but no big bang singularity at finite cosmic time}, should have the following three important ingredients:
\begin{enumerate}
 \item  An inflationary period at early times that could explain the origin of primordial inhomogeneities.

\item A sudden phase transition where particles are produced in a sufficient amount to reheat the universe at a viable temperature in order to match with the hot Friedmann universe.

\item An accelerated phase at late times, accounting for the current cosmic acceleration.
\end{enumerate}

Considering the above points, { and dealing with the flat Friedmann-Lema{\^\i}tre-Robertson-Walker (FLRW) spacetime}, to find such a background we start with a hydrodynamical fluid, the dynamics of which is given by a first order differential equation
of the Hubble parameter $H$ ($= \dot{a}/a$), of the form $\dot{H}= F(H)$, where $F$ is some continuous function of the Hubble parameter. For example, when one considers
a linear Equation of State (EoS) of the form $P(\rho)=(\gamma-1)\rho$ ({{}where $P$, $\rho$ are respectively, the pressure and the energy density of the
hydrodynamical fluid, and $\gamma \in [1,~2]$, is its Equation of State parameter}), one has $F(H)=-\frac{3\gamma}{2}H^2$.
One can easily verify that the solution of the first order differential equation for model $F(H)=-\alpha^2H^{\beta}$, with $\beta>1$ always predicts a finite { cosmic}
time singularity
at early times (i.e., the big bang singularity).
{{}Hence, in order to remove such singularity, for large values of the Hubble parameter, the function $F$ must be a linear function of the Hubble parameter.} {{}Secondly,}
to obtain a phase transition, we will assume that the derivative of $F$ is discontinuous at some point $H_E$, and we choose that, for $H\gtrsim H_E$,
$F(H)\cong -\frac{3\gamma}{2}H^2$ with $\frac{4}{3}<\gamma$. Due to the phase transition, the energy density of the produced $\chi$-particles, namely, $\rho_{\chi}$
decreases as $a^{-4}$, and the energy density of the background evolves as $a^{-3\gamma}$.
{{}Although,} the energy density of the produced particles is initially smaller than the energy density of the background, eventually the
density of the produced particles will be dominant and, as a result, the universe will be reheated, and hence it will evolve as the hot Friedmann universe.
{{}And finally,} if our model has to take into account the current cosmic acceleration, the simplest way is to assume that our dynamical system, {{}$\dot{H}= F(H)$, should
have a fixed point $H_f$} (hence, $F(H_f)=0$, i.e., a de Sitter solution), which can be modelled for $H_f\lesssim H\ll H_E$,
{{}by the function $F(H)=-\alpha^2 (H-H_f)^2$.}

An example of a nonlinear EoS (see the chapter 3 of \cite{Odintsov} for a review of cosmology with nonlinear EoS) that leads to a background with all the properties mentioned above,
is one that contains three parameters: a dimensionless one, namely $\gamma$,
and two energy densities {$\rho_f$, $\rho_e$ with the condition} $\rho_f\ll \rho_e$. Thus, it is given by
\begin{eqnarray}\label{EOS}
 P(\rho)=\left\{\begin{array}{ccc}
 -\rho+\gamma\sqrt{\rho_e}\left(2\sqrt{\rho}-\sqrt{\rho_e}\right)&\mbox{for}& \rho>\rho_E\\
    -\rho+ \gamma\left(\sqrt{\rho}-\sqrt{\rho_f}\right)^2            &\mbox{for}& \rho\leq\rho_E,
                \end{array}\right.
\end{eqnarray}
where
\begin{eqnarray}
 \rho_E=\left( \sqrt{\rho_e}+\sqrt{\rho_f}+\sqrt{2\sqrt{\rho_e\rho_f} }   \right)^2
\cong  {\rho_e}.
\end{eqnarray}

At energy densities $\rho\gg \rho_e$, the EoS of the model is approximately $P(\rho)\cong -\rho$, which mimics a cosmological constant,
leading to an accelerated phase at early times. {{}We will show that this late accelerating universe could have a viable inflationary period.}
After the phase transition, at scales $\rho= \rho_E$, the equation is $P(\rho)\cong (\gamma-1)\rho$, where if one chooses $\gamma=2$,
one has a stiff fluid that accounts for
a deflationary era.
{{}Finally, since the system has a unique fixed point, }
i.e., the equation $P(\rho)+\rho=0$, has only one solution, given by $\rho=\rho_f$, {hence,} at late times, the background asymptotically goes
to the de Sitter solution corresponding to this fixed point, and depicts the current cosmic acceleration.

The well-known Friedmann and Raychaudhuri equations in the flat FLRW geometry are
\begin{eqnarray}\label{Fried-Ray}
 H^2=\frac{\rho}{3}, \quad \dot{H}=-\frac{P(\rho)+\rho}{2},
\end{eqnarray}
and the combination of both leads to the {following} dynamical equation
\begin{eqnarray}
 \dot{H}=\left\{\begin{array}{ccc}
      -\frac{3\gamma}{2}H_e\left(2H-{H_e} \right)&\mbox{for}& H>H_E\\
       -\frac{3\gamma}{2}\left(H-H_f \right)^2&\mbox{for}& H\leq H_E,\\
                \end{array}\right.
\end{eqnarray}
where we have introduced the notation $H_i=\sqrt{\frac{\rho_i}{3}}$, for $i=e,~f$, and $H_E=\sqrt{\frac{\rho_E}{3}}=H_e+H_f+\sqrt{2H_e\,H_f}\cong {H_e}$.

The effective Equation of State (EoS) parameter, namely $w_{eff}$, which is defined as $w_{eff}=-1-\frac{2\dot{H}}{3H^2}$, for our model is given by
\begin{eqnarray}
 w_{eff}(H)=\left\{\begin{array}{ccc}
             -1+\gamma\frac{H_e}{H}\left(2-\frac{H_e}{H}  \right)&\mbox{for}& H>H_E\\
             -1+\gamma\left(1-\frac{H_f}{H}\right)^2&\mbox{for}& H\leq H_E,\\
            \end{array}\right.
\end{eqnarray}
which shows that for $H\gg H_e$ one has
$w_{eff}(H)\cong -1$ (early quasi de Sitter period). When $H_e\gtrsim H\gg H_f$, the EoS parameter satisfies
$w_{eff}(H)\cong -1+\gamma$, and finally, for $H\gtrsim H_f$ one also has  $w_{eff}(H)\cong -1$ (late quasi de Sitter period).

{This equation could  analytically be solved as}
\begin{eqnarray}\label{nonsingular}
 H(t)=\left\{\begin{array}{ccc}
  \frac{H_e}{2}\left[ \left(1+\frac{2H_f}{H_e}+\sqrt{\frac{8H_f}{H_e}}\right)e^{-{3\gamma}H_e\,t}+1                      \right]
  &\mbox{for}& t<0\\
  \frac{H_e+\sqrt{2H_eH_f}}{\frac{3\gamma}{2}(H_e+\sqrt{2H_eH_f})t+1}+H_f&\mbox{for}& t\geq 0,
             \end{array}\right.
\end{eqnarray}
which depicts  {{} a background without a finite cosmic time singularity in the past},
satisfying $H(-\infty)=\infty$, and $H(\infty)=H_f$, takes the following approximate form
\begin{eqnarray}
 H(t)\cong\left\{\begin{array}{ccc}
  \frac{H_e}{2} \left(e^{-{3\gamma}H_e\,t}+ 1\right)   &\mbox{for}& t<0\\
  \frac{H_e}{\frac{3\gamma}{2}H_e\,t+1}&\mbox{for}& t\geq 0.
             \end{array}\right.
\end{eqnarray}

\begin{remark}
 From the solution {of} $H(t)$, we will see that
 \begin{eqnarray}
  w_{eff}(t)\cong \left\{\begin{array}{ccc}
  -1+e^{3\gamma H_e t}& \mbox{for}& t\ll -\frac{1}{H_e}\\
  -1+\gamma & \mbox{for}& 0 \leq t\ll \frac{1}{H_f}\\
  -1+\frac{4}{3\gamma H_f^2\,t^2} & \mbox{for}&  t\gg \frac{1}{H_f}.
                         \end{array}\right.
 \end{eqnarray}

\end{remark}

Integrating $H$ one gets the following scale factor
\begin{eqnarray}\label{A}
a(t)=\left\{\begin{array}{ccc}
a_Ee^{-\frac{1}{6\gamma}\left( 1+\frac{2H_f}{H_e}+\sqrt{\frac{8H_f}{H_e}}   \right)\left[e^{-{3\gamma}H_e\,t}-1  \right]} e^{\frac{H_e}{2}t}&\mbox{for}& t<0
\\
a_E\left(\frac{3\gamma}{2}( H_e+\sqrt{2H_eH_f}   )t+1  \right)^{\frac{2}{3\gamma}}e^{{H_f}t}&\mbox{for}& t\geq 0,
\end{array}
\right.
\end{eqnarray}
which could be approximated by
\begin{eqnarray}
a(t)\cong\left\{\begin{array}{ccc}
a_Ee^{-\frac{1}{6\gamma}\left[e^{-{3\gamma}H_e\,t}-1  \right]}e^{\frac{H_e}{2}t}&\mbox{for}& t<0
\\
a_E\left(\frac{3\gamma}{2}H_e\,t+1 \right)^{\frac{2}{3\gamma}}&\mbox{for}& t\geq 0.
\end{array}
\right.
\end{eqnarray}

\vspace{0.5cm}

{{}
Before finishing this section,
an important remark about the incompleteness of our model is in order:
Note that the Hubble parameter, and thus the energy density, only diverge when $t\rightarrow -\infty$. This means that the big bang singularity
(understood as a divergence of the energy density at finite early cosmic time) is not present in this model. Actually, in analogy with the so-called {\it little  rip} singularity where the EoS parameter tends asymptotically to $-1$ at future time (see, for instance, \cite{fls,beno}), we may argue
that, in our model, the universe starts with a {\it little bang}. Moreover,
 following the arguments given in \cite{bgv}, we will see that our background
is not past-complete. This can be easily realized because for $t<0$, the scale factor in our model (\ref{A}) is given by
\begin{eqnarray}
a (t)= a_Ee^{-\frac{1}{6\gamma}\left( 1+\frac{2H_f}{H_e}+\sqrt{\frac{8H_f}{H_e}}   \right)\left[e^{-{3\gamma}H_e\,t}-1  \right]} e^{\frac{H_e}{2}t},
\end{eqnarray}
and thus,  the maximum affine parameter
$
 \tilde{\lambda}_{max}\equiv \frac{1}{a_E}\int_{-\infty}^0 a(t)dt
$ is finite, meaning that,  any backward-going null geodesic  has a finite affine length, i.e., it is past-incomplete.
The same happens with  massive particles moving along time-like geodesics. In this case, let $p_0\not= 0$ be the three-momentum  at time $t=0$, then  the maximum proper time
$\tau_{max}\equiv \int_{-\infty}^0 \frac{ma(t)}{\sqrt{m^2a^2(t)+p_0^2a^2(0)}}dt$ will also be finite, meaning that, in this system of reference, the singularity (the divergence
of the energy density) is at finite proper time.

In the same way, by choosing periodic potentials one can find a universe starting and ending in a de Sitter phase (see Eqs.~(26) and (68) of \cite{phpj}). The background actually leads to
a universe (although not geodesically past-complete)  where the energy density  never diverges.

Finally, note that  this analysis is only at the classical level, while for energy densities at Planck scales the classical picture losses it sense. In fact, from
the best of our knowledge, in the flat FLRW geometry, the only way to have a nonsingular universe which is geodesically complete is in the context of bouncing cosmologies \cite{bounces},
where in order to obtain a bounce one needs to introduce
nonstandard matter fields \cite{nonstandard} or to go beyond General Relativity \cite{LQC}.
}

\section{The scalar field}
\label{scalar-field}

At the early time, our background (\ref{nonsingular}) satisfies $P(\rho)\cong -\rho$, which indicates that the universe was of quasi de Sitter type, and
we aim to check whether this background could lead to a power spectrum of cosmological perturbations that fit well with the current observational data \cite{Ade}.

However,
it is unknown how the hydrodynamical
perturbations \cite{mfb}
could fit well with current observational data, because during the inflationary era, the square
of the velocity of sound $c_s^2$, appearing in the Mukhanov-Sasaki equation
\cite{ms}, becomes $c_{s}^{2}\equiv\frac{\dot{P}}{\dot{\rho}}\cong-1$ (remember that, we have $P(\rho)\cong -\rho$), and hence,
it leads to the Jeans instability for modes well inside the Hubble radius. But, in case of a universe driven by a single scalar field, one
always has $c_s^2= 1$, which means that we do not encounter any {such} instability in this case, and this is one of the main reasons to attempt to mimic
the dynamics governed by the hydrodynamical fluid with nonlinear equation of state described in section \ref{model} by a
single scalar field $\varphi$ with potential $V (\varphi)$.
{{} Nevertheless, the fundamental one, which is to find a physical justification over the choice of a  given nonlinear EoS is very complicated. In a similar way as in bulk viscosity or in gravitational induced ``adiabatic'' particle creations, we feel the same difficulties to justify some of the coefficients of viscosity and the particle creation rates used
	in the current literature \cite{pan}.} However, as we will immediately show, in our particular model, assuming a phase transition  to inflation to a deflationary regime \cite{spokoiny}, i.e., for $\gamma=2$,  which is what we will consider in the rest of the paper, we obtain an inflationary quintessential potential, the combination of the well-known $1$-dimensional Higgs potential with an exponential one.
{{}The first one is responsible for inflation and the exponential one leads,
    at late time, to the current cosmic acceleration. From our viewpoint, the simplicity of the obtained inflationary quintessential potential justifies the choice of the nonlinear EoS in our model.
}

\vspace{0.5cm}

{{} In order to find the potential, we bring in the scalar field dynamics. Hence, if we represent the energy density and the pressure of the scalar field by the notations $\rho_{\varphi}$, $p_{\varphi}$,
respectively, then in the flat FLRW background, they assume the following simplest forms}

\begin{align}\label{sf1}
\rho_{\varphi} & = \frac{1}{2} \dot{\rho}_{\varphi}^2+ V (\varphi),~~~p_{\varphi}= \frac{1}{2} \dot{\rho}_{\varphi}^2- V (\varphi)
\end{align}

Now, using Eq. (\ref{sf1}) and the Raychaudhuri equation in (\ref{Fried-Ray}), we find
\begin{eqnarray}
\varphi=\int\sqrt{-2\dot{H}}\,\,dt=-\int\sqrt{-\,\,\frac{2}{\dot{H}}}\,\,dH.
\end{eqnarray}
Now, in our case, for $H>H_E$, one gets
\begin{eqnarray}
\varphi=-2\sqrt{\frac{2}{3\gamma}}\sqrt{\frac{H}{H_e}-\frac{1}{2}}\Longleftrightarrow
H=\frac{H_e}{2}\left(\frac{3\gamma}{4}\varphi^2 +1\right),
\end{eqnarray}
and for $H\leq H_E$, one finds
\begin{eqnarray}
\varphi =\frac{-2}{\sqrt{3\gamma}}\ln\left(\frac{H-H_f}
{H_e+\sqrt{2H_eH_f}} \right)+\varphi_E\Longleftrightarrow H=\left(H_e+\sqrt{2H_eH_f}\right)e^{-\frac{\sqrt{3\gamma}}{2}(\varphi-\varphi_E )}+H_f,
\end{eqnarray}
where $\varphi_E\equiv -\sqrt{\frac{4}{3\gamma}}\sqrt{1+\frac{2H_f}{H_e}+\sqrt{\frac{8H_f}{H_e}}}\cong -\frac{2}{\sqrt{3\gamma}} $.

{{}Similarly, using Eq. (\ref{sf1}) and Eq. (\ref{Fried-Ray}), the potential  of the scalar field becomes}
$V(H)=3H^2+\dot{H}$. {{} For the case of a phase transition to a deflationary regime studied in \cite{spokoiny,pv} , i.e., for the choice $\gamma=2$, one has}
\begin{eqnarray}\label{potencialH} V(H)=\left\{\begin{array}{ccc}
3H^2-6HH_e+3H_e^2&\mbox{for}& H>H_E\\
6HH_f-3H_f^2&\mbox{for}& H\leq H_E.\end{array}\right.
\end{eqnarray}

That is,
\begin{eqnarray}\label{POT} V(\varphi)=\left\{\begin{array}{ccc}
\frac{27H_e^2}{16}\left(\varphi^2-\frac{2}{3}\right)^2
&\mbox{for}& \varphi<\varphi_E\\
3H_f^2\left[2\left(\frac{H_e}{H_f}+\sqrt{\frac{2H_e}{H_f}}\right)e^{-\sqrt{\frac{3}{2}}(\varphi-\varphi_E)}+1  \right]
&\mbox{for}& \varphi\geq \varphi_E.\end{array}\right.
\end{eqnarray}

{ To understand the behavior of our model potential (see Eq. (\ref{POT})), we display it in figure 1, which shows that the potential is a decreasing function of
the scalar field. The figure also indicates that, after a certain time, the potential becomes { nearly} zero. }

\begin{figure}[h]
\begin{center}
\includegraphics[scale=0.50]{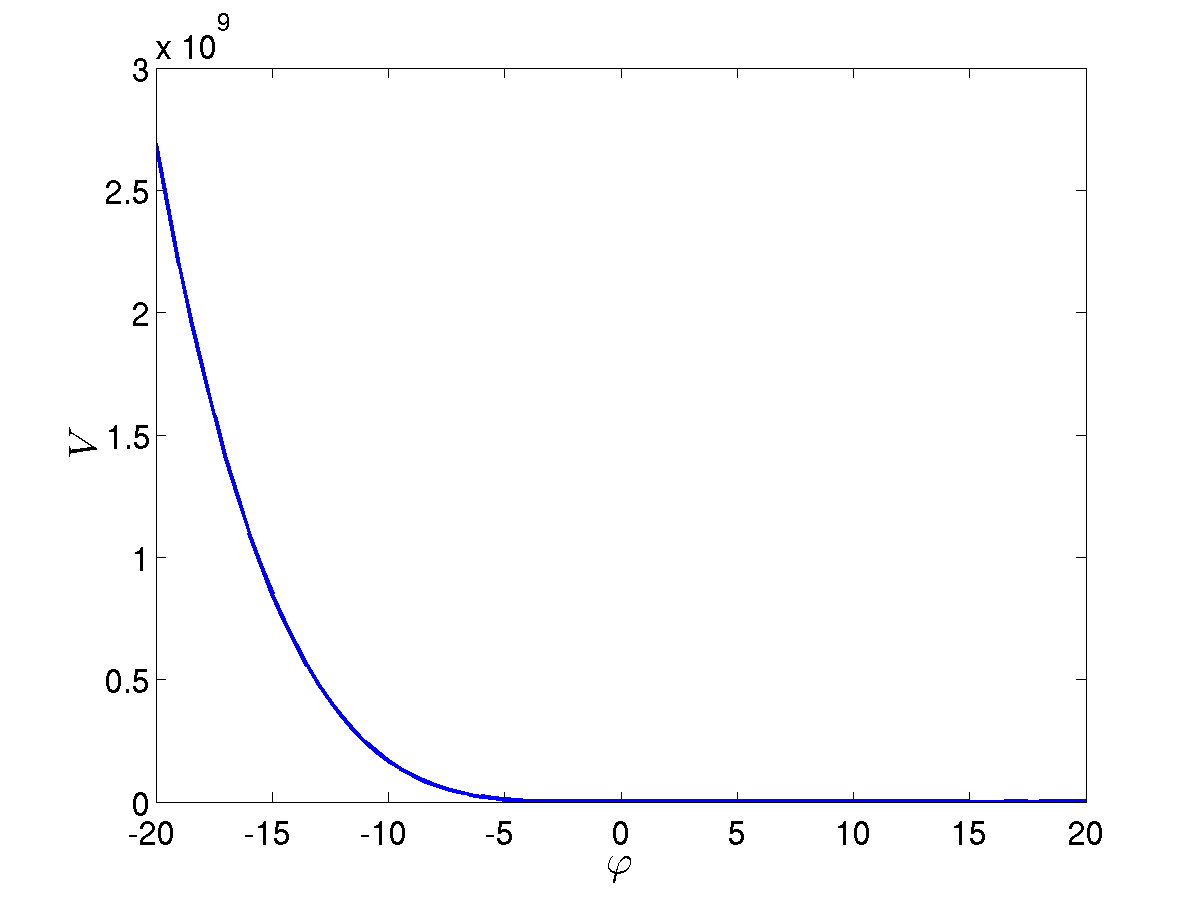}
\end{center}

\caption{{\protect\small The figure shows the shape of the potential for $H_e=10^2$, and $H_f=1$.}}
\end{figure}

{{} In fact we can see that it depicts the $1$-dimensional Higgs potential, or also called  Double Well Inflationary potential \cite{DWI} --
{\it it has the shape of a $1$-dimensional Mexican hat} for $\varphi<-\sqrt{\frac{2}{3}}$, and nearly vanishing, and thus,  having a deflationary regime, for
$\varphi\geq -\sqrt{\frac{2}{3}}$.}



{{} Once we have obtained our background, via reconstruction method},
the conservation equation now leads to the second order differential equation
\begin{eqnarray}\label{sf-conservation}
 \ddot{\varphi}+\sqrt{3}\sqrt{\frac{\dot{\varphi}^2}{2}+V(\varphi)}\,\,\,\dot{\varphi}+V_{\varphi}(\varphi)=0,
\end{eqnarray}
{where $V_{\varphi}(\varphi)$ is the differentiation with respect to $\varphi$,} {{} having infinitely many solutions which lead to different backgrounds}. One of
these solutions is that which leads to our background, i.e.,
\begin{eqnarray}
 \varphi(t)=\left\{\begin{array}{ccc}
                    -\sqrt{\frac{2}{3}}\sqrt{ 1+\frac{2H_f}{H_e}+\sqrt{\frac{8H_f}{H_e}}   }e^{-3H_et}& \mbox{for}& t<0\\
                    \sqrt{\frac{2}{3}}\ln\left(3\left(H_e+\sqrt{2H_eH_f}\right)t+1 \right)+\varphi_E& \mbox{for}& t\geq 0,
                   \end{array}\right.
\end{eqnarray}
{{} is a solution of (\ref{sf-conservation}), that replaced   in $H(t)=\frac{1}{\sqrt{3}}\sqrt{\frac{\dot{\varphi}^2(t)}{2}+V(\varphi(t))}$, leads to the
background (\ref{nonsingular}) with $\gamma=2$.}

\section{Cosmological perturbations}
Now, in order to study the cosmological perturbations, we need to study the slow roll parameters \cite{btw}
\begin{eqnarray}\label{slowroll}
 \epsilon=-\frac{\dot{H}}{H^2}, \quad \eta=2\epsilon-\frac{\dot{\epsilon}}{2H\epsilon},
\end{eqnarray}
which allows us to calculate the associated parameters, namely, the spectral index ($n_s$), its running ($\alpha_s$), and the ratio of tensor to scalar perturbations ($r$) defined as
\begin{eqnarray}
 n_s-1=-6\epsilon+2\eta, \quad \alpha_s=\frac{H\dot{n}_s}{H^2 +\dot{H}},\quad
 r=16\epsilon.
\end{eqnarray}
and finally, we can compare them with the current observational data to realize a viable inflationary potential.

Now, at early times, i.e., when $H>H_E$,
if we consider a phase transition to a deflationary era (i.e., $\gamma= 2$), {then introducing a new variable $x\equiv \frac{6H_e}{H}$, the slow roll parameters (\ref{slowroll}) can be expressed in an alternate way as }
\begin{eqnarray}
 \epsilon=x\left(1-\frac{x}{12}\right),\quad \eta=\epsilon+\frac{x}{2},
\end{eqnarray}
and as a consequence, we find $
n_s-1=-3x+\frac{x^2}{3}.$
Conversely,
\begin{eqnarray}
 x=\frac{9}{2}\left(1-\sqrt{1-\frac{4(1-n_s)}{27}}\right).
\end{eqnarray}

Now, from the observational bound on the spectral index, we can find its range. As the
recent PLANCK$+$WP 2013 data
(see table $5$ of \cite{Ade}) predicts the spectral index at 1$\sigma$ Confidence Level (C.L.) to be $n_s=0.9583\pm 0.0081$, this means that, at $2\sigma$ C.L., one has $0.0085\leq x \leq 0.0193$,
and thus, $0.1344\leq r=16\epsilon\leq 0.3072$.

Since  PLANCK$+$WP 2013 data also provides the constrain $r\leq 0.25$, at $95.5\%$ C.L., thus,
when $0.0085\leq x \leq 0.0156$,
the spectral index belongs to the  1-dimensional
marginalized $95.5\%$ C.L., and also $r\leq 0.25$, at $95.5\%$ C.L.

Further, for the running at 1$\sigma$ C.L., PLANCK+WP 2013 data gives $\alpha_s=-0.021\pm 0.012$, and our background (\ref{nonsingular}) leads to the theoretical value $\alpha_s\cong -\frac{3x\epsilon}{1-\epsilon}\cong -3x^2$. Consequently, at the scales we are dealing with,  $-7\times 10^{-4}\leq \alpha_s\leq -2\times 10^{-4}$, and thus, the running also belongs to the  1-dimensional marginalized $95.5\%$ C.L.

{{} Furthermore, as the effective equation of state is given by $w_{eff}(H)=-1+\frac{2}{3}\epsilon$,
 and if we assume $\epsilon= 1$, denotes the end of the inflation where $H_{end}$ be the value of the Hubble parameter, then
$w_{eff}(H_{end})=-\frac{1}{3}$, which implies that the universe will just enter into the decelerating phase. }

{{}On the other hand, if $N(H)$ denotes the number of e-folds from observable scales exiting the Hubble radius, then it can be calculated by using
$N(H)=-\int_{H_{end}}^H\frac{H}{\dot{H}}dH$, leading to the following equation}

\begin{eqnarray}
N(x)=\frac{1}{x}-\frac{1}{x_{end}}+\frac{1}{12}\ln\left(\frac{12-x}{12-x_{end}}\frac{x_{end}}{x} \right),
\end{eqnarray}
where $x_{end}=6(1-\sqrt{2/3})\cong 1.1010$, denotes the value of the parameter $x$, when inflation ends. For the values of $x$ that allow to fit well  with the theoretical value of the spectral index, its running and the tensor/scalar ratio with their corresponding observable values,
we will obtain $64\leq N(x)\leq 117$.

\vspace{0,5cm}
 In the same way we have considered the 2-dimensional marginalized confidence level in the plane $(n_s,r)$ in the presence of running (see figure $2$), where
 the black path corresponds to curve $((n_s(x),r(x))$ provided by our model.

{{}We notice a slight difference (or, updated measurements) between Planck 2013 and Planck 2015 constraints as follows:
when one considers Planck2013 data \cite{Ade} at $95\%$ C.L., 
the parameter $x$ ranges in the interval $[0.0075, 0.0156]$ which means that the number of e-folds is constrained to $64\leq N(x)\leq 133$. On the other hand, if one considers Planck2015 TT$+$low P data \cite{Planck} at $95\%$ C.L., the parameter $x$ ranges in
the interval $[0.0061, 0.0124]$, and hence, $80\leq N(x)\leq 163$.}

\begin{figure}
\includegraphics[height=0.37\textwidth,angle=0]{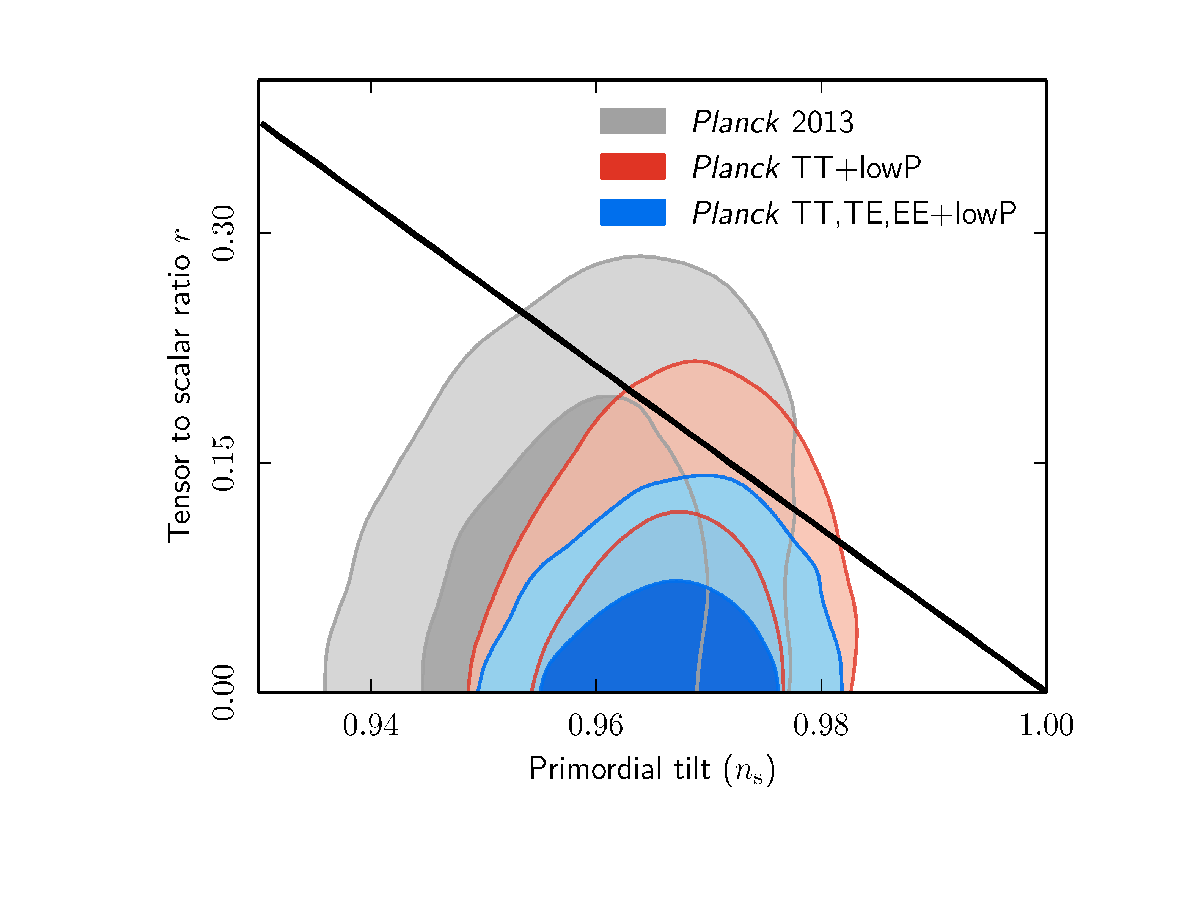}
\caption{Marginalized joint confidence contours for $(n_{\mathrm s} \,, r)$,
at the 68\,\% and 95\,\% CL, in the presence of running of the spectral indices. The black path corresponds to the curve $(n_s(x),r(x)))$ of our model
({Figure courtesy of the Planck2015 Collaboration}).
}
\label{fig:nsrr}
\end{figure}

\vspace{0,5cm}

To determinate the value of $H_e$, one has to take into account the theoretical \cite{btw} and the observational \cite{bld} value of the power spectrum
\begin{eqnarray}\label{power1}
 {\mathcal P}\cong \frac{H^2}{8\pi^2\epsilon}=
 \frac{36H_e^2}{\pi m^2_{pl}\epsilon x^2}\cong 2\times 10^{-9},
\end{eqnarray}
where we have  explicitly introduced the Planck's mass,
which in our units takes $m_{pl}=\sqrt{8\pi}$. {Finally,} using the values of $x$ in the range
{$[0.0075,0.0156]$,}
we may conclude that
\begin{eqnarray} H_e\sim  10^{-8}{m_{pl}}\sim 10^{11} \mbox{ GeV}.
\end{eqnarray}

\section{The reheating process}

In this section, we will study the reheating of the universe via gravitational particle production due to a sudden phase transition. First of all, we need to define the modes that depict
the vacuum state before and after the phase transition.  The Klein-Gordon equation corresponding to the
the Lagrangian density of a scalar field is \cite{Birrell} ${\mathcal
L}=\frac{1}{2}(\partial_{\mu}\phi\partial^{\mu}\phi -m^2\phi^2-\xi
R\phi^2)$, and its  corresponding  Klein-Gordon equation is given by
\begin{eqnarray}\label{aaa1}
 (-\nabla_{\mu}\nabla^ {\mu}+ m^2+\xi R)\phi=0,
\end{eqnarray}
where $\xi$ is the coupling constant, and $R=6(\dot{H}+2H^2)$, is the scalar
curvature.
The modes of the form $\phi_{\bf k}({\bf
x},\tau)\equiv \frac{e^{i{\bf kx}}}{(2\pi)^{-\frac{3}{2}}a(\tau)} \chi_{\bf
k}(\tau)$ ($\tau$ is the conformal time) will satisfy the equation
\begin{eqnarray}\label{a2}
\chi''_{k}(\tau)+\Omega_{ k}^2(\tau)\chi_{ k}(\tau)=0,
\end{eqnarray}
where we have introduced the notation $\Omega_{ k}^2(\eta)\equiv
\omega_{k}^2(\tau)+(\xi-1/6)a^2(\tau)R(\tau)$, with $\omega_{
k}^2(\eta)=m^2a^2(\tau)+k^2$.

The definition of the vacuum modes, will depend on the particles involved in the production process, and their coupling to gravity. We will start with the
the production of massless particles which are nearly conformally coupled to gravity, and
just for mathematical simplicity, we will choose, in our model,  $H_f=0$, and then $H_E={H_e}$. As a result,
after the phase transition, our universe exactly went through a deflationary phase, on consideration $\gamma=2$.

Then, at early and late time the term $a^2R$ will vanish, and the modes (exact solutions of (\ref{a2})) given by
\begin{eqnarray}\label{a36}
 \chi_{in, k}(\tau)=\frac{e^{-ik\tau}}{\sqrt{2{k}}}-\frac{\xi-1/6}{{k}}
\int_{-\infty}^{\tau}a^2(\tau)R(\tau)\sin({ k}(\tau-\tau'))\chi_{k}(\tau')d\tau',\\
\chi_{out, k}(\tau)=\frac{e^{-ik\tau}}{\sqrt{2{k}}}+\frac{\xi-1/6}{{k}}
\int_{\tau}^{\infty}a^2(\tau)R(\tau)\sin({ k}(\tau-\tau'))\chi_{k}(\tau')d\tau',
\end{eqnarray}
have the following behavior at early and late times respectively
\begin{eqnarray}
 \chi_{in, k}(\tau)\simeq \frac{e^{-ik\tau}}{\sqrt{2{k}}} (\mbox{ when }\tau\rightarrow -\infty), \quad \chi_{out, k}(\tau)\simeq \frac{e^{-ik\tau}}{\sqrt{2{k}}}
 (\mbox{ when } \tau\rightarrow +\infty).
\end{eqnarray}

Since we are considering particles nearly conformally coupled {to gravity}, we can consider the term $(\xi-1/6)a^2(\tau)R(\tau)$ as a perturbation, and we can approximate the ``in'' and ``out'' modes
by the first order Picard's iteration {as}
\begin{eqnarray}\label{a37}
 \chi_{in, k}(\tau)\cong \frac{e^{-ik\tau}}{\sqrt{2{k}}}-\frac{\xi-1/6}{{k}\sqrt{2{k}}}
\int_{-\infty}^{\tau}a^2(\tau)R(\tau)\sin({ k}(\tau-\tau')) e^{-ik\tau'} d\tau',\\
\chi_{out, k}(\tau)\cong \frac{e^{-ik\tau}}{\sqrt{2{k}}}+\frac{\xi-1/6}{{k}\sqrt{2{k}}}
\int_{\tau}^{\infty}a^2(\tau)R(\tau)\sin({ k}(\tau-\tau'))e^{-ik\tau'}d\tau',
\end{eqnarray}
which will represent, respectively, the vacuum before and after the phase transition.

Then after the phase transition, we could write the ``in'' mode as a linear combination of the ``out'' mode and its conjugate as follows

\begin{eqnarray}
 \chi_{in,k}(\tau)=\alpha_k\chi_{out,k}(\tau)+\beta_k\chi^*_{out,k}(\tau),
\end{eqnarray}
where $\alpha_k$ and $\beta_k$ are the so-called Bogoliubov coefficients. Imposing the continuity of $\chi$ and its first derivative at the transition time we obtain,
up to order $\left(\xi-1/6  \right)^2 $,
the value of these
coefficients:
\begin{eqnarray}\label{bogoliubov}
  \alpha_k\cong 1-\frac{i({\xi}-\frac{1}{6})}{2k}\int_{-\infty}^{\infty}a^2(\tau)
 R(\tau) d\tau,\quad  \beta_k\cong \frac{i({\xi}-\frac{1}{6})}{2k}\int_{-\infty}^{\infty}e^{-2ik\tau}a^2(\tau)
 R(\tau) d\tau.
\end{eqnarray}

Finally,
the energy density of the produced particles due to the phase transition is given by
\cite{Birrell}
\begin{eqnarray}
 \rho_{\chi}=\frac{1}{2\pi^2a^4}\int_0^{\infty}k^3|\beta_k|^2 dk,
\end{eqnarray}

The integral of the $\beta$-Bogoliubov coefficient (\ref{bogoliubov}) is convergent because at early and late time, the term $a^2(\tau)R(\tau)$ converges fast enough to zero.
Moreover, applying integration by parts twice on the $\beta$-Bogoliubov coefficient (\ref{bogoliubov}), it can be shown that, it becomes
$\beta_k\sim {\mathcal O}(k^{-3})$ (this is due to the fact that $\dot{H}$ is continuous during the phase transition)
{{}and we will see,
this implies that} the energy density of produced particles is not ultra-violet divergent. Moreover,
 $\beta_k=({\xi}-\frac{1}{6})f(\frac{k}{a_EH_e})$, where $f$ is some function,
then
the energy density of the produced particles approximately becomes
\begin{eqnarray}
\hspace{-1cm} \rho_{\chi} \cong \left({\xi}-\frac{1}{6}\right)^2{\mathcal N}H_e^{4}\left(\frac{a_E}{a}\right)^4,
\end{eqnarray}
where we have introduced the notation
$${\mathcal N}\equiv \frac{1}{2\pi^2}\int_0^{\infty}s^3f^2(s)ds.$$

\begin{remark}
 We have bounded this quantity. For the upper bound, we have used the Cauchy-Schwarz inequality and the Plancherel theorem to obtain

 \begin{eqnarray*}{\mathcal N}\leq \frac{1}{32\pi H_e^4a_E^4}\sqrt{\int_{-\infty}^{\infty}a(t) \left(\frac{d(a^2(t) R(t))}{dt}\right)^2dt} \,\,\,\,\,
 	\sqrt{\int_{-\infty}^{\infty}a^3(t) R^2(t)dt}\,\,\,=\\
 	\frac{1}{32\pi}\sqrt{\frac{3}{2}\int_{1}^{\infty}\frac{e^{-\frac{5}{12}(y-1)}}{y^{\frac{7}{12}}}(1-21y-15y^2+y^3)^2dy+\frac{576}{10}}\,\,\,\,\,
 	\sqrt{\frac{3}{2}\int_1^{\infty}\frac{e^{-\frac{1}{4}(y-1)}}{y^{\frac{5}{2}}}(y^2-4y+1)^2dy+6}\,\,\,\cong \, 9.5 \mbox{ }.
 \end{eqnarray*}

 To obtain the lower bound note that
 $${\mathcal N}\geq \frac{1}{2\pi^2 H_e^4a_E^4({\xi}-\frac{1}{6})^2}\int_{\Lambda a_EH_e}^{\infty}k^3|\beta_k|^2 dk,
 $$
 where $\Lambda$ is some positive dimensionless constant.

 On the other hand, we have approximated the value of the $\beta$-Bogoliubov coefficient, {by performing three times integration by parts and retaining the leading term as}
 $$\beta_k\cong \frac{81}{4k^4}H_e^4a_E^4\left({\xi}-\frac{1}{6}\right),
 $$
 and thus, choosing $\Lambda \cong 30$ in order that the leading term will be dominant, we obtain
 $${\mathcal N}\geq \frac{6561}{128\pi^2 \Lambda^4}\cong  \frac{5}{\Lambda^4}\cong 6\times 10^{-6}.
 $$

\end{remark}

Since the sudden transition occurs at
${H_e}\sim 10^{-8} m_{pl}\sim 10^{11}$ GeV,
and at the transition time, the energy density of the produced particles is of the order
\begin{eqnarray}
 \rho_{\chi}\sim   10^{-36}\left({\xi}-\frac{1}{6} \right)^{2}{\mathcal N} \rho_{pl},
 \end{eqnarray}
where $\rho_{pl}=m_{pl}^4$ is the Planck's energy density. Note that, at the transition time, this energy density
is smaller than the energy density of the background which is of the order $\rho\sim \frac{3}{8\pi}H_e^2m_{pl}^2 \sim  10^{-17} \rho_{pl}.$

Now, after this phase transition,
the produced particles
interact with each other exchanging gauge bosons between them, and
constitute a relativistic plasma that thermalises the universe \cite{spokoiny,pv} before it was radiation
dominated. {{}Moreover, as the background in our model
went through a deflationary phase, that means its energy density} decays as $a^{-6}$
and  the energy density of the produced particles  decreases as $a^{-4}$.
Naturally, the energy density of the produced particles will dominate, and the
universe will become radiation dominated, which matches with the standard hot Friedmann universe.
The universe will expand, subsequently it will gradually cool, and, as a result the produced
particles will be non-relativistic and, thus, the universe will enter a matter dominated phase, which is
essential for the cosmological perturbations to grow. Further,
only at very late time, when the
energy density of the background is of the same  order of $\rho_1$, the field comes back,
and we observe the current accelerating universe.

The reheating temperature, namely, $T_R$,
is defined as the temperature of the universe {{}when both the energy densities of the background and of the produced particles are of the same order}
 ($\rho\sim \rho_{\chi}$).
Since {{} $\rho_{\chi}\cong ({\xi}-\frac{1}{6})^2{\mathcal N} H_e^4\left(\frac{a_E}{a}\right)^4$, and $\rho\cong H_e^2 m_{pl}^2\left(\frac{a_E}{a}\right)^6$, one obtains,
 $\frac{a_E}{a(t_R)}\sim \left|{\xi}-\frac{1}{6}\right|\sqrt{{\mathcal N}}\,\,\frac{H_e}{m_{pl}}$, and therefore, we find the reheating temperature as
\begin{eqnarray}\label{reheating-temp}
T_R\sim \rho_{\chi}^{1/4}(t_R)\sim  \left|{\xi}-\frac{1}{6}\right|^{\frac{3}{2}}{\mathcal N}^{\frac{3}{4}}\frac{H_e^2}{m_{pl}}\sim 10^3
\left|{\xi}-\frac{1}{6}\right|^{\frac{3}{2}} {\mathcal N}^{\frac{3}{4}}\mbox{ GeV }.
\end{eqnarray}
}
{{}The reheating temperature in Eq. (\ref{reheating-temp})} is below the GUT scale $10^{16}$ GeV, which means that the GUT symmetries are not restored
preventing a second  monopole production stage. Moreover, it guarantees the standard successes with nucleosynthesis, as it  requires a reheating temperature
below $10^{9}$ GeV \cite{37}.

Lastly, we aim to find the temperature of the universe when an equilibrium is reached. We follow the thermalization process as described by the authors in Refs. \cite{pv,spokoiny}, where
it has been assumed that the interactions between the produced particles are due to exchange of gauge bosons. The interaction rate, $\Gamma$, is given by $\Gamma=n_{\chi}\sigma$,
where $\sigma$, the cross section of scattering, is
$\sigma\sim \frac{\alpha^2}{\bar{\epsilon}^2}$ ($\alpha$ is
a coupling constant, and $\bar{\epsilon}\sim H_e\left(\frac{a_E}{a} \right)$ denotes the typical energy of a produced particle), and the density of produced
particles is \cite{Birrell, Haro}

\begin{eqnarray}
  n_{\chi}=\frac{1}{2\pi^2 a^3}\int_0^{\infty}k^2|\beta_k|^2 dk
  =\left({\xi}-\frac{1}{6}\right)^2{\mathcal M}H_e^{3}\left(\frac{a_E}{a}\right)^3,
\end{eqnarray}
 where, for our model, we find{{}
\begin{eqnarray}{\mathcal M}\equiv \frac{1}{16\pi a_E^3H_e^3}\int_{-\infty}^{\infty} a^4(\tau)R^2(\tau) d\tau
=\frac{3}{32\pi}\int_1^{\infty}\frac{e^{-\frac{1}{4}(y-1)}}{y^{\frac{5}{2}}}(y^2-4y+1)^2dy+\frac{3}{8\pi}
\simeq 7 \times 10^{-1}.
\end{eqnarray}

{{} Since the thermal equilibrium is achieved when $\Gamma\sim H(t_{eq})=H_e\left(\frac{a_E}{a_{eq}}\right)^3$ (recall that, in our model, this process is produced in
the deflationary phase, that means the energy density evolves as $a^{-6}$),
also, one has
$\frac{a_E}{a_{eq}}\sim \alpha |{\xi}-\frac{1}{6}|{\mathcal M}^{\frac{1}{2}}$, hence,
the equilibrium temperature becomes $T_{eq}\sim \left|{\xi}-\frac{1}{6} \right|^{\frac{5}{4}}{\mathcal N}^{\frac{1}{4}}{\mathcal M}^{\frac{1}{2}}\alpha H_e$.
Therefore, using the value of $H_e$, one obtains
\begin{eqnarray}
T_{eq}\sim 10^{11}\left|{\xi}-\frac{1}{6} \right|^{\frac{5}{4}}{\mathcal N}^{\frac{1}{4}}{\mathcal M}^{\frac{1}{2}}\alpha \mbox{ GeV}.
\end{eqnarray}

And choosing as usual $\alpha\sim (10^{-2}-10^{-1})$
\cite{spokoiny, pv}, {{}one finds that the equilibrium temperature lies in the following range
\begin{eqnarray}
T_{eq}\sim \left(10^9-10^{10}\right)\left|{\xi}-\frac{1}{6} \right|^{\frac{5}{4}}{\mathcal N}^{\frac{1}{4}} {\mathcal M}^{\frac{1}{2}}\mbox{ GeV }.
\end{eqnarray}
}
}

\vspace{1cm}

To end this section, we will study the production of heavy massive particles ($m\gg H_e$) conformally coupled to
gravity due to a phase transition to a deflationary regime (see \cite{he} for details). In that case,
since the Hubble parameter is continuous up to second derivative,
during the adiabatic regimes,
we will use the second order WBK solution to define approximately the vacuum modes \cite{Haro}
\begin{eqnarray}
\chi_{2,k}^{WKB}(\tau)\equiv
\sqrt{\frac{1}{2W_{2,k}(\tau)}}e^{-{i}\int^{\tau}W_{2,k}(\eta)d\eta},
\end{eqnarray}
where $W_{2,k}$ was calculated in \cite{Bunch} {{}as}
\begin{eqnarray}
W_{2,k}=
\omega_k-\frac{m^2a^4}{4\omega_k^3}(\dot{H}+3H^2)+\frac{5m^4a^6}{8\omega_k^5}H^2+
\frac{m^2a^6}{16\omega_k^5}(\dddot{H}+15\ddot{H}H+10\dot{H}^2+86\dot{H}H^2+60H^4)\\
-\frac{m^4a^8}{32\omega_k^7}(28\ddot{H}H+19\dot{H}^2+394\dot{H}H^2+507H^4)+\frac{221m^6a^{10}}{32\omega_k^9}(\dot{H}+3H^2)H^2-\frac{1105m^8a^{12}}{128\omega_k^{11}}H^4.
\end{eqnarray}

Then
the square modulus of the $\beta$-Bogoliubov coefficient
is obtained by calculating the square modulus of the Wronskian ${\mathcal W}[\chi_{2,k}^{WKB}(0^-), \chi_{2,k}^{WKB}(0^+)]$ (see \cite{Haro,he}), where
$\chi_{2,k}^{WKB}(\tau)$ is evaluated after and before the phase transition. It is not difficult to show that it
will be given by
\begin{eqnarray}
 |\beta_k|^2\cong \frac{m^4a_E^{12}\left(\dddot{H}(0^+)-\dddot{H}(0^-)\right)^2}{1024(k^2+m^2a^2_E)^6},
\end{eqnarray}
where $\dddot{H}(0^-)$ (resp. $\dddot{H}(0^+)$), is the value of the second derivative of the Hubble parameter before (after) the phase transition. Recall that, for our model,
this quantity is discontinuous at the transition epoch.

Then the number density of produced particles and their energy density will respectively be
\begin{eqnarray}
 n_{\chi}\sim 10^{-2}\frac{H_e^8}{m^5}\left(\frac{a_E}{a} \right)^3, \quad \rho_{\chi}\sim mn_{\chi}.
\end{eqnarray}

These particles  are far from being in thermal equilibrium. The more massive particles will decay into lighter
particles, which will interact through multiple scattering, thus  a re-distribution of energies among the different
particles occurs -- {\it kinetic equilibrium}, and also, an increase in the number of particles
-- {\it chemical equilibrium}. That is, in such process, in order to obtain a relativistic
plasma in thermal equilibrium, both number--conservating and number--violating reactions are definitely involved \cite{37}.

To calculate when thermalization occurs, we consider as in \cite{37}, {{}which tells that} the thermalization rate of these processes is $\Gamma=\alpha^2 n_{\chi}^{\frac{1}{3}}$. Then
the equilibrium
is reached when $\Gamma\sim H(t_{eq})=H_e\left(\frac{a_E}{a_{eq}}\right)^3$, obtaining
$\frac{a_E}{a_{eq}}\sim 2\alpha \left(\frac{H_e}{m}\right)^{\frac{5}{6}}$. {Finally, at the time of equilibrium, the energy densities of both produced particles and the background
respectively take the forms}
\begin{eqnarray}
 \rho_{\chi}(t_{eq})\sim 8\times 10^{-2}\alpha^3{H_e^4}\left(\frac{H_e}{m}\ \right)^{\frac{13}{2}}, \quad \rho(t_{eq})\sim 8\alpha^6 H_e^2 m_{pl}^2\left(\frac{H_e}{m}\ \right)^{{5}}.
\end{eqnarray}

After this thermalization, the relativistic plasma evolves as $\rho_{\chi}(t)=\rho_{\chi}(t_{eq})\left(\frac{a_E}{a} \right)^4$, and the background evolves as
$\rho(t)=\rho(t_{eq})\left(\frac{a_E}{a} \right)^6$, because we are in the deflationary regime. The reheating is obtained when both energy densities are of the same order, {{}and that} will happen when $\frac{a_E}{a_{eq}}\sim \sqrt{\frac{\rho_{\chi}(t_{eq})}{\rho(t_{eq})}}$, {{}and thus, we
obtain the reheating temperature of the order}
\begin{eqnarray}
 T_R\sim \rho_{\chi}^{\frac{1}{4}}(t_{eq})\sqrt{\frac{\rho_{\chi}(t_{eq})}{\rho(t_{eq})}}\sim \alpha^{-\frac{3}{4}}\frac{H_e^2}{m_{pl}}\left(\frac{H_e}{m}\
 \right)^{\frac{19}{8}}
 \sim 10^4\left(\frac{H_e}{m}\ \right)^{\frac{19}{8}} \mbox{ GeV}.
\end{eqnarray}

\section{COMPARISON WITH OTHER QUINTESSENTIAL MODELS}
\label{other-potentials}

In Ref. \cite{spokoiny}, the author proposed some models to depict quintessential inflation, some of them have the following forms:
\begin{eqnarray}\label{Vphi}
 V(\varphi)=V_0\left(\frac{e^{-\lambda_1 \varphi}}{1+e^{\lambda_2 \varphi}}\right),\quad \mbox{or,}\,\,\,\,V(\varphi)=V_0e^{-\lambda \varphi^2},
\end{eqnarray}
where $\lambda_1, \lambda_2$ and $\lambda$ are positive parameters.

The main problem of those models is that the potential was too smooth to produce a sufficient amount of particles to reheat the universe.
As the number of produced particles
is an adiabatic invariant \cite{parker}, so for smooth potentials, this quantity is negligible. Only an abrupt phase transition that breaks down the adiabaticity, which can be mimicked by a discontinuity of some
derivative of the potential, could produce a high amount of particles.

One of such possible potentials,
proposed by Peebles and Vilenkin in \cite{pv} follows
\begin{eqnarray}\label{potential}
 V(\varphi)=\left\{\begin{array}{ccc}
   \lambda(\varphi^4+M^4)&\mbox{for}& \varphi\leq 0\\
   \frac{\lambda M^8}{\varphi^4+ M^4}&\mbox{for}& \varphi\geq  0,
                   \end{array}\right.
\end{eqnarray}
where  $\lambda$ and $M$ are {simply real} parameters. In \cite{pv}, the authors chose $\lambda=10^{-14}$, and $M=8\times 10^{-14} m_{pl}$.

The model leads to a power spectrum that matches at $95.5 \%$ C.L. with Planck2013 data. Effectively, it is easy to see that
\begin{eqnarray}
\epsilon\cong \frac{8}{\varphi^2}, \quad \eta\cong \frac{12}{\eta^2},
\end{eqnarray}
which means that the spectral index and the tensor to scalar ratio are related by $1-n_s=\frac{3}{16}r$ (The same relation happens for our model because in our case we have $\epsilon\cong x$, and thus, $1-n_s\cong 3x$, and $r\cong 16x$). Then, using Planck2013 data, one can check that, for $0.136\leq r\leq 0.25$, both parameters belong to the
$1$-dimensional marginalized $95.5\%$ C.L. Moreover, the number of e-folds is given by $N(r)\cong \frac{16}{r}$, then for the range of viable values of $r$, one has $ 64\leq N(r)\leq 117$, which coincides with the same range of e-folds that we have obtained for our model.

{ To obtain the value of $\lambda$, one has to use the formula (\ref{power1}), {where for that model, we have} $H^2\cong \frac{V}{3}\cong \frac{\lambda \varphi^4}{3}$,
and $r\cong \frac{128}{\varphi^2}$. {Thus, one has}
\begin{eqnarray}{\mathcal P}\sim \frac{8^5\lambda }{3\pi^2r^3}\sim 2\times 10^{-9}\Longrightarrow \lambda\sim \frac{6\pi^2r^3}{8^5}\times 10^{-9}
\Longrightarrow 4\times 10^{-15}\leq \lambda \leq 3\times 10^{-14}.
\end{eqnarray}

And to calculate the value of the parameter $M$, the authors realize that for inflationary quintessential models, after reheating, the scalar field starts to dominate only when the
potential energy at the reheating time is of the same order {with} the present energy density $\rho_0=3 m_{pl}H_0^2$ (where $H_0\sim 10^{-60} m_{pl}$ is the current value of the
Hubble parameter). Then, since at the reheating time, one has $\varphi(t_R)\sim m_{pl}$, the potential energy will be $V(\varphi(t_R))\sim \lambda M^8 m_{pl}^{-4}$, and thus,
\begin{eqnarray}
M\sim \lambda^{-1/8}m_{pl}^{3/4}H_0^{1/4}\sim 10^{-13} m_{pl}\sim 10^6 \mbox {GeV }.
\end{eqnarray}

\begin{remark}
 From this viewpoint, 
 {for our model, when} the reheating is due to the production of massless conformally coupled particles, {then using formulas (\ref{potencialH}) and
 (\ref{reheating-temp}), we obtain}
 {{}
 \begin{eqnarray}
  H_f\sim \frac{m_{pl}^3H_0^2}{H_e^4}{\mathcal N}^{-\frac{3}{2}}\left|\xi-\frac{1}{6}\right|^{-3}\sim 10^{-88}{\mathcal N}^{-\frac{3}{2}}\left|\xi-\frac{1}{6}\right|^{-3} m_{pl}.
 \end{eqnarray}
 }

 And, when the reheating is done via heavy particles conformally coupled {to} gravity, we have
 \begin{eqnarray}
  H_f\sim 10^3\frac{H_0^2m_{pl}^3}{H_e^4}\left(\frac{m}{H_e} \right)^{\frac{9}{4}}\sim 10^{-85}\left(\frac{m}{H_e} \right)^{\frac{9}{4}} m_{pl}.
 \end{eqnarray}

\end{remark}
}

\vspace{0.5cm}

However, the  model proposed by Peebles and Vilenkin contains two obscure points that were not clarified in their paper:
\begin{enumerate}
 \item  
 {The authors claim (based on the investigations by Ford \cite{ford})
 that,} immediately after the end of inflation, the universe enters into a deflationary regime. However,
 what is proved in \cite{ford} is that for a potential with an absolute minimum (this is not our case), and a shape of the form $V(\varphi)=\lambda \varphi^{2n}$, when the field
 oscillates,
 the effective EoS of the background is $P=\frac{n-1}{n+1}\rho $, and thus, its energy density decreases as $a^{-\frac{6n}{n+1}}$, and as a consequence, it decays faster
 than radiation for $n>2$.
 {And for the inflationary quintessential potentials (non-oscillatory models), the effective EoS, $P=\rho$ (deflationary regime) is reached only at very late times
 	(that means when $\varphi\rightarrow +\infty$) for the potentials, that approach a constant
 	value.}


\item The authors claim (based in the results obtained in \cite{Damour-Vilenkin} and \cite{Giovannini}) that 
the energy density of produced massless minimally coupled particles is
\begin{eqnarray}\label{vilenkin}
\rho_{\chi}={\mathcal R} H_E^4\left(\frac{a_E}{a} \right)^4,\end{eqnarray}
where ${\mathcal R}\sim 10^{-2}$.  However, in those works, in order to perform {an analytic computation},  it is assumed that the transition from the
de Sitter phase to another with a constant effective EoS parameter
is too abrupt. In fact, the Hubble parameter is
discontinuous at the transition time, and thus, the energy density of the produced particles is ultra-violet divergent. {For} this reason, the authors take a frequency cut-off to calculate
$\rho_{\chi}$, and state that, if the transition was smoother, this quantity will be convergent, and the value of ${\mathcal R}$ will be of
the same order as the value obtained by Ford in \cite{ford}, where the author considers that the creation of massless particles nearly conformally coupled to gravity is due to a
smooth phase transition to a radiation {{} or matter} dominated universe ($\dot{H}$ was continuous),
which of course, is model
dependent,
and it is not clear at all whether it applies to the model (\ref{potential}). In fact, for this potential, up to fourth derivative, the potential
is continuous  what means that up to fifth derivative the Hubble parameter is continuous, and thus,  the particle
production will be lower than in our model (where the third derivative of the Hubble parameter is discontinuous at the transition time), and as a final consequence, maybe leading to an
abnormally low reheating temperature.
\end{enumerate}

The first point is  essential for the viability of the model  because if one is dealing with light minimally coupled particles, in order to calculate mode solutions of equation
(\ref{a2}), and thus, to define the vacuum state after
the phase transition
one needs a  nearly constant effective EoS parameter ($w_{eff}$). In that case, the vacuum modes  after the phase transition will be (see \cite{he}, where
 a similar
situation was studied in the context of  bouncing cosmologies)
\begin{eqnarray}\label{a48}
 \chi_{out, k}(\tau)\cong e^{-i(\frac{\pi\nu_{out}}{2}+\frac{\pi}{4})}
 \sqrt{\frac{\pi\tau}{4}}H^{(2)}_{\nu_{out}}({ k} \tau),
\end{eqnarray}
where $H^{(2)}_{\nu}$ is the  Bessel's functions with
$\nu_{out}\equiv\sqrt{\frac{1}{4}+\frac{2(1-3w_{eff})}{(1+3w_{eff})^2}}$.

On the other hand, although $w_{eff}$ was constant after the phase transition,  when the field is not
nearly conformally coupled, it would be a very difficult task to find vacuum modes before the phase
transition, because to obtain them one firstly needs to calculate numerically the scale factor as a function of the conformal time,
which could only be done solving numerically the conservation equation obtaining  $\varphi(t)$, and consequently the Hubble parameter $H(t)$. With this Hubble parameter,
one can numerically calculate the scale factor as a function of the conformal time,
and finally, once the scale factor has been calculated, one has
to solve the differential equation (\ref{a2}).

\vspace{1cm}

Coming back to the model, we perform
a change of variable $\varphi=M\bar\varphi$, and by re-parameterizing the time coordinate as $t=\frac{1}{M\sqrt{\lambda}}\bar{t}$, the conservation equation
(\ref{sf-conservation}) takes the simple form
\begin{eqnarray}\label{equation}
 \bar{\varphi}''+\sqrt{3}M\sqrt{\frac{(\bar{\varphi}')^2}{2}+
 \bar{V}(\bar\varphi)}\,\,\,\bar{\varphi}'+\bar{V}_{\bar\varphi}(\bar\varphi)=0,
\end{eqnarray}
where `\,$'$\,' denotes the derivative with respect to the {{}new time coordinate} $\bar t$, {{}and in terms of the new coordinates, the potential in (\ref{potential}) becomes}
\begin{eqnarray}\label{XXX}
 \bar{V}(\bar\varphi)=\left\{\begin{array}{ccc}
   \bar{\varphi}^4+1&\mbox{for}& \bar{\varphi}\leq 0\\
   \frac{1}{\bar{\varphi}^4+ 1}&\mbox{for}& \bar{\varphi}\geq  0.
                   \end{array}\right.
\end{eqnarray}

In the new variables the effective EoS parameter is given by
\begin{eqnarray}\label{efectiva}
w_{eff}(\bar t)= -1+\frac{(\bar{\varphi}')^2}{\frac{(\bar{\varphi}')^2}{2}+\bar{V}(\bar\varphi)  }.\end{eqnarray}

We want to integrate the system (\ref{equation}), taking as initial condition when the inflation ends. This happens when the parameter $\epsilon=-\frac{\dot{H}}{H^2}$ is of the
order one. Taking into account that $w_{eff}=-1+\frac{2}{3}\epsilon$, we obtain from (\ref{efectiva}) that inflation ends when $\bar{\varphi}'_{end}=\sqrt{\bar{\varphi}_{end}^4+1}$,
where $\bar{\varphi}_{end} $ is the value of the field at the end of inflation. To obtain that value we will use that during slow roll $\epsilon$ is approximately given by
$\epsilon=\frac{1}{2}\left(\frac{V_{\varphi}}{V}\right)^2$, then the condition $\epsilon\cong 1$ leads to the value $\varphi_{end}\cong 2\sqrt{2}$. Finally, using that the variables
$\varphi$ and $\bar\varphi$ are related by the relation $\varphi=M\bar\varphi$, we will obtain that the initial condition to integrate numerically the system (\ref{XXX}) will be
\begin{eqnarray}
 \left(\bar\varphi_0=\frac{2\sqrt{2}}{M}, \bar\varphi_0'=\sqrt{\frac{64}{M^4}+1}\right).
\end{eqnarray}

{
We have integrated the dynamical system (\ref{equation}) for $M=1$, and $M=0.25$  (for smaller values of $M$, we have checked numerically {and found} that the behavior 
{remains} same as {it is for} $M=0.25$). We see in figure $3$ that,
after the phase transition, a deflationary regime does not exist, rather it seems that $w_{eff}\cong -0.8$, and in that case, the energy density of the created particles will
decay faster than
the one of the background, which means that the universe never reheats. On the other hand, for a smaller value of $M$ (we have considered $M=0.25$), after the end of the inflation,
the universe immediately enters  into the deflationary regime (see figure $4$). In that case, eventually the energy density of the created particles will dominate, and the universe will be
reheated. Then, for small enough $M$, it seems that the numerical calculations show  that  the first claim of Peebles and Vilenkin is right. However, regarding the value of { ${\mathcal R}$}, we do not
know how to calculate approximately the ``in'' modes. Consequently, their second claim  remains an open question for us.

\begin{figure}[h]
\begin{center}
\includegraphics[scale=0.30]{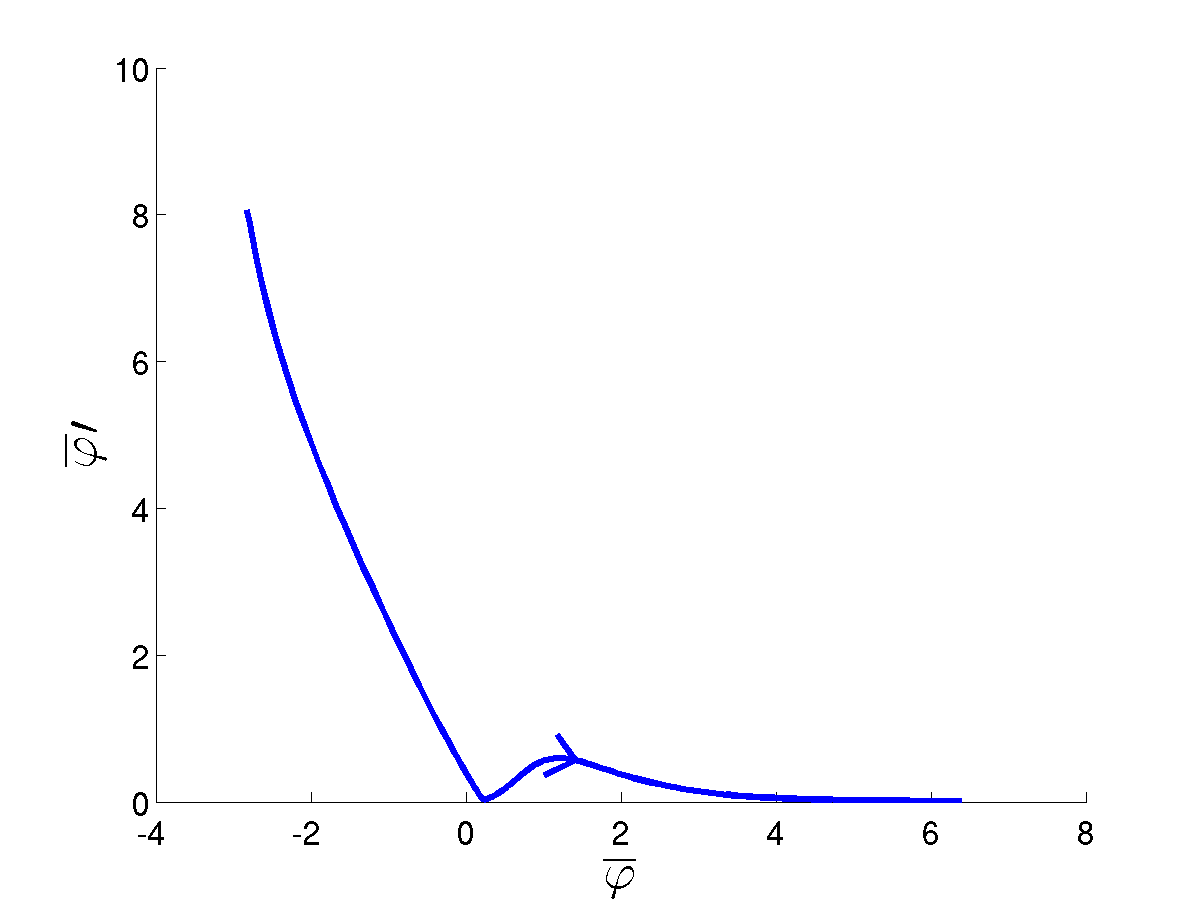}
\includegraphics[scale=0.30]{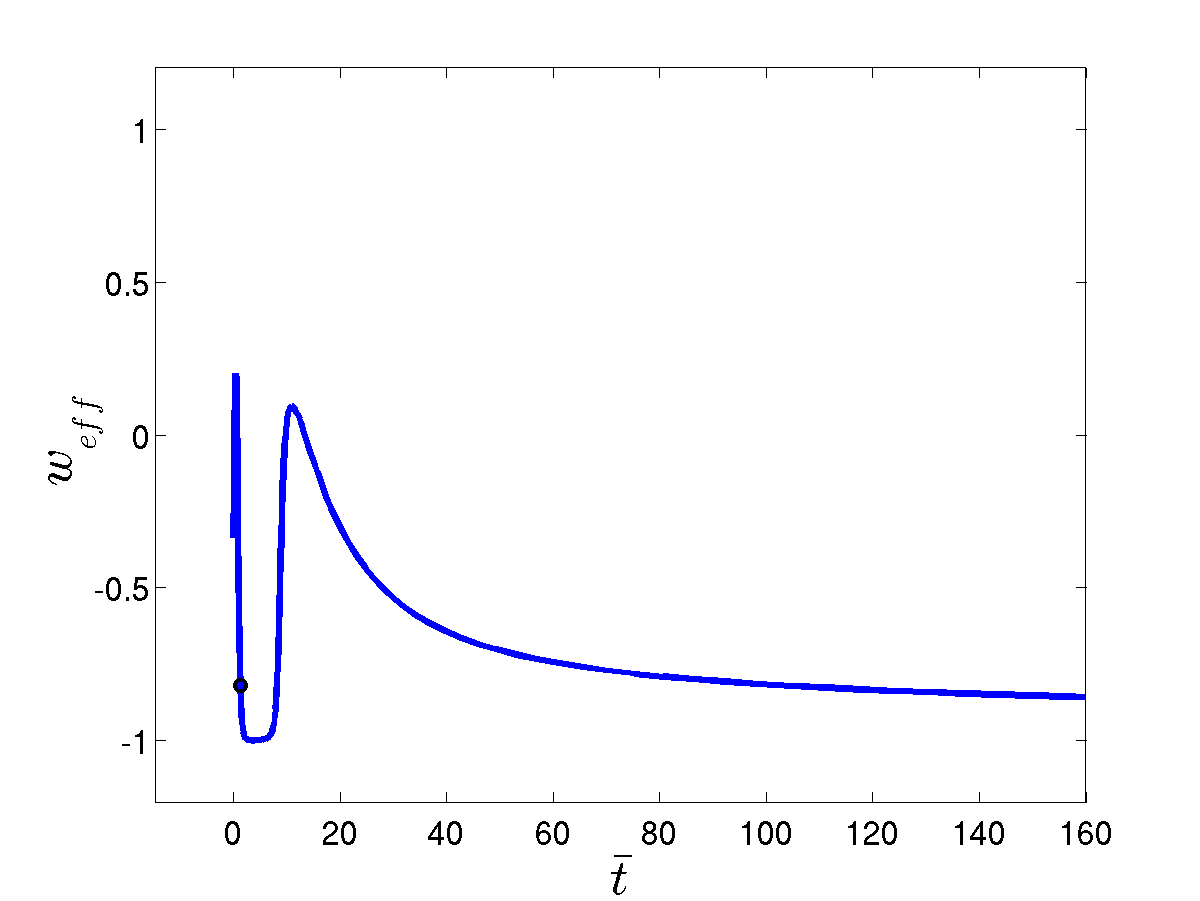}
\end{center}

\caption{{\protect\small Phase portrait and $w_{eff}$ for $M=1$. The black point in the second picture represents when the phase transition occurs. }}
\end{figure}

\begin{figure}[h]
\begin{center}
\includegraphics[scale=0.30]{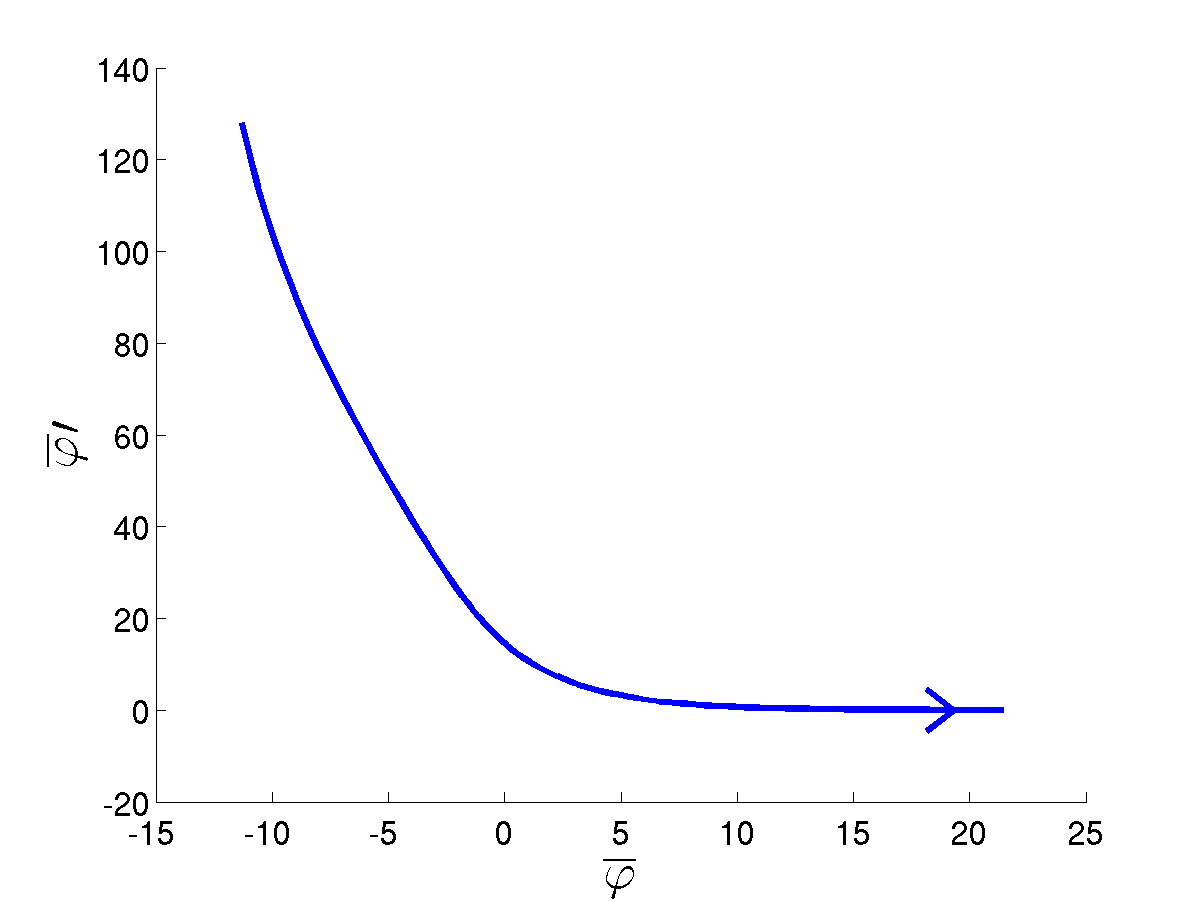}
\includegraphics[scale=0.30]{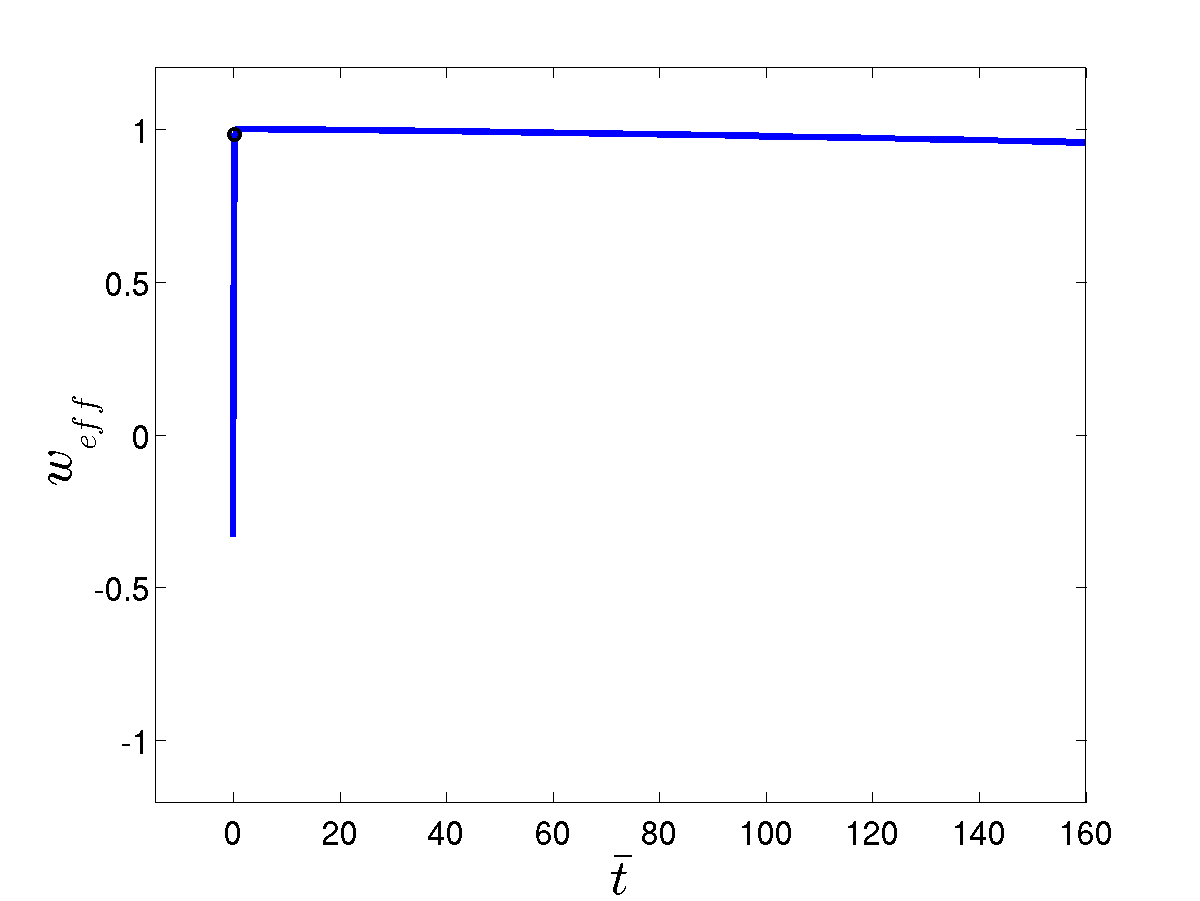}
\end{center}

\caption{{\protect\small Phase portrait and $w_{eff}$ for $M=0.25$. The black point in the second picture represents when the phase transition occurs.}}
\end{figure}

}

{

\section{The simplest model}
{Now, we come to the last section, where} 
we consider the case
 of a cosmological constant, namely, $\Lambda$, different from zero.  Performing the replacement $\rho\rightarrow \rho+\Lambda$ in the  Friedmann equation, this  will become
$H^2=\frac{\rho +\Lambda}{3}$. In the same way, choosing $\gamma=2$, and performing the replacement
$\rho\rightarrow \rho+\Lambda$, $\rho_e\rightarrow \rho_e+\Lambda$ and $\rho_f\rightarrow 0$ in the EoS
(\ref{EOS}) we obtain the background as
\begin{eqnarray}\label{bbb}
H(t)=\frac{1}{\sqrt{3}}\left\{\begin{array}{ccc}
\frac{1}{3}\left((2M+3\sqrt{\Lambda})e^{-\frac{8}{\sqrt{3}}Mt}+2M\right)& \mbox{for} & t<0\\
& & \\
\sqrt{\Lambda}\left(  \frac{2M+3\sqrt{\Lambda}+2Me^{-2\sqrt{3\Lambda}t}}{2M+3\sqrt{\Lambda}-2Me^{-2\sqrt{3\Lambda}t}}         \right)            & \mbox{for} & t>0,
\end{array}\right.
\end{eqnarray}
where once again $M=\frac{3\sqrt{3}}{4}H_e $.

For this background, the effective EoS parameter satisfies all  requirements needed to depict a viable
universe:
\begin{eqnarray}
w_{eff}=\left\{\begin{array}{ccc}
-1 & \mbox{for} & t\ll -\frac{1}{M}\\
1 &\mbox{for} & 0<t\ll\frac{1}{\sqrt{\Lambda}}\\
-1 &\mbox{for} & t\gg\frac{1}{\sqrt{\Lambda}}.
\end{array}\right.
\end{eqnarray}

And using the reconstruction method, 
{the potential takes the  following simplest form}
\begin{eqnarray}\label{simple}
V(\varphi)=\left\{\begin{array}{ccc}
M^2\left(\varphi^2-\frac{2}{3}\right)^2& \mbox{for} & \varphi<\varphi_E\\
\Lambda& \mbox{for} & \varphi\geq \varphi_E,
\end{array}\right.
\end{eqnarray}
where $\varphi_E=\sqrt{\frac{\sqrt{\Lambda}}{M}+\frac{3}{2}}$. Note that, {{} at finite cosmic time, the ``nonsingular''} background
(\ref{bbb})
 comes from this potential, and allows us to perform analytical calculations, such as, the spectral index, the ratio of tensor to scalar perturbations, and the reheating temperature provided by the model.

Recalling that, in General Relativity, the potential that mimics a hydrodynamical fluid with linear EoS
$P=(\gamma-1)\rho$ is given by
\begin{eqnarray}
\bar{V}(\varphi)=\left\{\begin{array}{ccc}
|V_0|e^{-\sqrt{3\gamma}\varphi}& \mbox{for}& \gamma<1\\
0& \mbox{for}& \gamma=1\\
-|V_0|e^{-\sqrt{3\gamma}\varphi}& \mbox{for}& \gamma>1.
\end{array}\right.
\end{eqnarray}
We will see that our simple potential (\ref{simple}), is the matching of a double well inflationary potential (which depicts the inflationary phase), with a potential that mimics a
stiff fluid (which depicts the deflationary regime) plus a cosmological constant (which depicts the current cosmic acceleration).

Finally, note that,
all the results obtained above  {can be applied to this model}, that is, the theoretical values about {the} cosmological
perturbations match correctly with Planck's data, and also, it leads to a consistent reheating with recent observations.
The values of our parameters are {easily} calculated: due to the relation between $H_e$ and $M$, one has
$M\sim 10^{-8}m_{pl}$, and to calculate the value of the cosmological constant, one has to use the fact that the cosmological constant will be dominant when it will be of the
same order as the present energy density, that is, when $\Lambda\sim H_0^2m^2_{pl}\sim 10^{-120}\rho_{pl}$.

}

\section{Summary of the work}
\label{discuss}

{{} The essence of this work is to realize, although geodesically past incomplete,  but a universe without the big bang singularity at finite cosmic time, that
unifies the inflation with late time acceleration.}
{{}Considering a hydrodynamical fluid with nonlinear equation of state, we found a scenario of our universe}
which predicts an early inflationary phase of quasi de Sitter type, then
a sudden phase transition from inflation to a deflationary era,
and finally,  the current accelerating universe of de Sitter type. {Just after this inflationary phase, the universe immediately entered into a stiff matter era,
following a sudden phase transition, which as a result, produces an enough amount of particles. After this particle productions, the energy density of the produced particles
gradually became dominant
in compared to the background energy density}. Hence, the universe is reheated.
After that, the model goes to the de Sitter
universe in an asymptotic manner showing the current accelerating universe. {\textit{Thus, the present model can be considered as a unified  cosmic model
{{}without the big bang singularity}}.}

{{} It has been shown that the early inflationary
stage obtained by our nonlinear EoS can be mimicked by an equivalent field theoretic description which provides 
 the well-known $1$-dimensional Higgs potential, and since we have an analytic solution (\ref{nonsingular}),
 it is possible to calculate
the analytic expressions for the cosmological parameters, such as, the spectral index,
its running, and
the ratio of tensor to scalar perturbations, which, as we have showed, are all in good agreement with the Planck 2013 observations (see \cite{Ade}).}


{We have also} compared our model potential with the existing inflationary quintessential
potentials \cite{spokoiny,pv}, and show that our model is an improved version of the models \cite{spokoiny,pv} due to having its analytic behavior
which do not exist in the above models. {Finally, introducing a cosmological constant, we have seen that our model simplifies a lot, but it maintains
the same kind of properties, {present in} our previous quintessential model.
}


\vspace{0.5cm}

{\it Acknowledgments.--}
This investigation has been supported in part by MINECO (Spain), projects MTM2014-52402-C3-1-P and MTM2012-38122-C03-01. SP is supported in part by CSIR, Govt. of India.

\end{document}